\documentclass[aps,pre,10pt,groupedaddress,superscriptaddress,showkeys,nofootinbib,twocolumn]{revtex4-2}

\usepackage[driverfallback=dvipdfm]{hyperref}
\usepackage{amsmath,amssymb}
\usepackage{mathtools}
\usepackage{fancyvrb}

\allowdisplaybreaks[1]

\usepackage{color}

\newcommand{\Tr}{\mathrm{Tr}}
\newcommand{\tr}{\mathrm{tr}}

\newcommand{\drm}{\mathrm{d}}

\newcommand{\atan}{\mathrm{atan}}
\renewcommand{\vec}[1]{\boldsymbol{#1}}
\newcommand{\HQET}{\mathrm{HQET}}
\newcommand{\imp}{\mathrm{imp}}

\renewcommand{\vec}[1]{\boldsymbol{#1}}

\begin{document}


\title{QCD Kondo effect for single heavy quark in chiral-symmetry broken phase}


\author{Shigehiro~Yasui}
\email[]{yasuis@keio.jp}
\affiliation{International Institute for Sustainability with Knotted Chiral Meta Matter (WPI-SKCM$^{2}$), Hiroshima University, Higashi-Hiroshima, Hiroshima 739-8526, Japan}
\affiliation{Research and Education Center for Natural Sciences, Keio University, Hiyoshi 4-1-1, Yokohama, Kanagawa 223-8521, Japan}
\author{Daiki~Suenaga}
\email[]{daiki.suenaga@kmi.nagoya-u.ac.jp}
\affiliation{Kobayashi-Maskawa Institute for the Origin of Particles and the Universe, Nagoya University, Nagoya 464-8602, Japan}
\affiliation{Research Center for Nuclear Physics, Osaka University, Ibaraki 567-0048, Japan}
\author{Kei~Suzuki}
\email[]{k.suzuki.2010@th.phys.titech.ac.jp}
\affiliation{Advanced Science Research Center, Japan Atomic Energy Agency (JAEA), Tokai 319-1195, Japan}


\date{\today}

\begin{abstract}
We consider the QCD Kondo effect for a single heavy quark in quark matter composed of light quarks with chiral symmetry breaking.
Introducing several spinor structures in
QCD Kondo condensates, i.e., particle-projected condensate, antiparticle-projected condensate, and normal condensate without projection, we calculate the 
attractive energy gained by
the heavy quark
within the mean-field approximation
in the path-integral formalism.
We show that the normal condensate is favored at low density and the particle-projected condensate is favored at high density, when the light quark has a nonzero mass.
We interpret such a density-dependent transition between the two condensates in terms of the Kondo resonances.
\end{abstract}


\maketitle

\section{Introduction}
\label{sec:introduction}

Heavy quarks with the charm or bottom flavor are important for researching the strong interaction governed by Quantum Chromodynamics (QCD).
In quark matter, the heavy quarks are regarded as impurity particles in the medium composed of light quarks, such as up and down quarks, and they can lead to the Kondo effect at low temperature~\cite{Yasui:2013xr,Hattori:2015hka}.
This is called the QCD Kondo effect.
It is caused by the gluon exchange between a light quark and a heavy quark.
Indeed, the color exchange with the $\mathrm{SU}(N_{c})$ symmetry ($N_{c}=3$ is the number of colors) enhances the strength of the effective interaction between quarks at low-energy scale.
This mimics the original Kondo effect which is caused by the spin exchange with the $\mathrm{SU}(2)$ symmetry for impurity atoms in metals~\cite{Kondo:1964}, see, e.g., Refs.~\cite{Hewson,Yosida,Yamada,coleman_2015} for more details.\footnote{One may regard mathematically the QCD Kondo effect as an extension from the $\mathrm{SU}(2)$ symmetry to the $\mathrm{SU}(N)$ symmetry. Furthermore, it is possible to realize the Kondo effect in various non-Abelian groups, see Ref.~\cite{Kimura:2020uhq}.}
The Kondo effects are realized also in Dirac/Weyl metals and semimetals~\cite{Principi:2015,Yanagisawa:2015conf,Yanagisawa:2015,Mitchell:2015,Sun:2015,Feng:2016,Kanazawa:2016ihl,Lai:2018,Ok:2017,PhysRevB.97.045148,PhysRevB.98.075110,Dzsaber:2018,PhysRevB.99.115109,KIM2019236,Grefe:2019,Grefe:2020,Pedrosa:2021,Silva:2024kpj}.
Thus the Kondo effect can be a universal phenomenon
from condensed matter physics to high-energy physics.

The QCD Kondo effect was first studied in the perturbation theory by the Nambu--Jona-Lasinio model~\cite{Yasui:2013xr} and the one-gluon exchange model~\cite{Hattori:2015hka}.
In both cases, it was concluded that the perturbation is not applicable at a low-energy scale, called the Kondo scale.
This is because the coupling strength becomes effectively larger due to the quantum effect: the virtual creation and annihilation of the particles and holes near the Fermi surface.
Thus, the Kondo effect inevitably induces nonperturbative quantum phenomena.

The non-perturbative analysis of the QCD Kondo effect was conducted in the mean-field theory under the assumption that heavy quarks are distributed uniformly in three-dimensional space ~\cite{Yasui:2016svc,Yasui:2017izi}.\footnote{In the condensed matter physics, since the early days, the mean-field approximation has been adopted to the analysis of the Kondo effect for a single impurity~\cite{Takano:1966,Yoshimori:1970,ReadNewns1983,Eto:2001,Yanagisawa:2015conf,Yanagisawa:2015}. See also Ref.~\cite{Hewson} and references therein.}
The mean field was introduced to indicate an expectation value for the composite field, which mixes the light quarks and the heavy quarks in the ground state.
This is called the QCD Kondo condensate.
It has interesting topological properties, e.g., the Berry phases and the monopoles, in the momentum space~\cite{Yasui:2017izi}.
The QCD Kondo condensate was explored also for a single heavy quark, where the QCD Kondo condensate appears as the resonance state (QCD Kondo cloud)
 near the Fermi surface in the quark matter made of massless quarks~\cite{Yasui:2016yet}.

The purpose of the present study is to explore the QCD Kondo effect
for a {\it single} heavy quark in matter made of {\it massive} light quarks.
So far the QCD Kondo effect for a single heavy quark was studied for the matter made of {\it massless} quarks~\cite{Yasui:2016yet}, but not for the {\it massive} one.
However, the matter composed of massive light quarks, where the chiral symmetry for light quarks is broken, is more realistic than the massless one, because the light quarks should have current and/or dynamical masses.
So far, the influence of the chiral symmetry breaking for light quarks was investigated for heavy quarks that are {\it distributed uniformly} in three-dimensional space~\cite{Suzuki:2017gde}, but not for an {\it isolated single} heavy quark.
However, a single heavy quark may be more realistic in relativistic heavy-ion collisions, because produced heavy quarks would be a small number, and the
heavy quarks may not be regarded
to be distributed uniformly.
Thus, the present work fills the gaps in the previous studies.

This paper is organized as follows.
In Sec.~\ref{eq:formalism}, we introduce the
Lagrangian for the QCD Kondo effect for a single heavy quark in matter made of massive light quarks.
We introduce several types of the QCD Kondo condensates
and adopt the mean-field approximation to calculate the impurity energy decrease caused by the QCD Kondo condensate.
In Sec.~\ref{sec:results}, we present the numerical results for the impurity energy,
and
give interpretations 
in terms of the QCD Kondo resonances.
The final section is devoted to the conclusion and outlooks.

\section{Formalism}
\label{eq:formalism}

\subsection{Lagrangian for heavy and light quarks}
\label{sec:model_lagrangian}

We consider the situation that a single heavy quark exists as an isolated impurity in matter whose quark mass is $m$.
We suppose one flavor for the light quarks.
Based on the
procedure in Ref.~\cite{Yasui:2016yet},
we introduce the partition function
\begin{eqnarray}
   Z
=
   \int {\cal D}\psi {\cal D}\bar{\psi} \, {\cal D}\hat{f} {\cal D}\hat{f}^{\dag} {\cal D}\lambda \,
   \exp\biggl( \int \mathrm{d}^{4}x {\cal L}\bigl[\psi,\hat{f},\lambda\bigr] \biggr),
\label{eq:partition_function_0}
\end{eqnarray}
where the
Lagrangian is given by
\begin{eqnarray}
   {\cal L}\bigl[\psi,\hat{f},\lambda\bigr]
&= &
 - \bar{\psi} \bigl( \gamma\partial + m - \mu\gamma_{4} \bigr) \psi
\nonumber \\ && 
+ G
   \Bigl(
         \bar{\psi} \hat{f} \, \hat{f}^{\dag} \psi
      + \bar{\psi} i\gamma_{5} \hat{f} \, \hat{f}^{\dag} i\gamma_{5} \psi
         \nonumber \\ && \hspace{2em} 
      + \bar{\psi} i\vec{\gamma} \hat{f} \, \hat{f}^{\dag} i\vec{\gamma} \psi
      + \bar{\psi} i\vec{\gamma}\gamma_{5} \hat{f} \, \hat{f}^{\dag} i\vec{\gamma}\gamma_{5} \psi
   \Bigr)
   \delta(\vec{x})
   \nonumber \\ && 
 - \hat{f}^{\dag} \partial_{\tau} \hat{f} \, \delta(\vec{x})
 - \lambda \bigl( \hat{f}^{\dag}\hat{f} - 1 \bigr) \, \delta(\vec{x}),
\label{eq:Lagrangian_fermion_LH_isolated_2}
\end{eqnarray}
with the coordinate
$x^{\mu}=(\vec{x},\tau)$
 in the four-dimensional Euclidean
space\footnote{In the Euclidean spacetime,
we use the following representations of the Dirac matrices given by
\begin{eqnarray}
   \gamma_{k}
=
\left(
\begin{array}{cc}
 0 & -i\sigma^{k} \\
 i\sigma^{k} & 0  
\end{array}
\right), \,
   \gamma_{4}
=
\left(
\begin{array}{cc}
 I & 0 \\
 0 & -I
\end{array}
\right), \,
   \gamma_{5}
=
\left(
\begin{array}{cc}
 0 & -I \\
 -I & 0  
\end{array}
\right), \label{GammaDef}
\end{eqnarray}
with the Pauli matrices $\sigma^{k}$ ($k=1,2,3$) and the $2\times2$ dimensional unit matrix $I$.
}
and the chemical potential $\mu$ for the light quarks.
In Eq.~\eqref{eq:Lagrangian_fermion_LH_isolated_2}, $\psi(x)$ is the relativistic Dirac spinor
for the light quark with Dirac mass $m$, and $\hat{f}(\tau)$ is the nonrelativistic
 spinor for the heavy quark.
We notice that $\hat{f}(\tau)$ has dependence only on the Euclidean time $\tau$ because the heavy quark is supposed to 
locate
at the origin of three-dimensional space, $\vec{x}=\vec{0}$.
This setting is expressed by the three-dimensional $\delta$-function, $\delta(\vec{x})$.
The auxiliary field $\lambda(\tau)$ is the Lagrange multiplier for the constraint condition, $\hat{f}^{\dag}(\tau)\hat{f}(\tau) = 1$, indicating that the number of heavy quarks should be one.

The second term in Eq.~\eqref{eq:Lagrangian_fermion_LH_isolated_2} represents
the interaction between the light and heavy quarks provided by the point-like interaction with the scalar, pseudoscalar, vector, and axialvector channels with the common coupling constant $G$~\cite{Yasui:2016svc,Yasui:2017izi}.
This mimics a one-gluon exchange interaction.
The Lagrangian
 can be obtained
from the viewpoint of the heavy quark effective theory, see Appendix~\ref{sec:HQET_Lagrangian}.

We adopt the mean-field approximation for the four-point interaction in Eq.~\eqref{eq:Lagrangian_fermion_LH_isolated_2}.
The mean field gives the QCD Kondo condensate, i.e., the dynamical mixing between the light quark and the heavy quark in the ground state.
To proceed with the calculation,
 we introduce the auxiliary fields: the scalar field $\Phi(x)$, the pseudoscalar field $\Phi_{5}(x)$, the vector field $\vec{\Phi}(x)$, and the axialvector field $\vec{\Phi}_{5}(x)$.
These functions correspond to the heavy-light bifermion forms $\bar{\psi}(x)\Gamma\hat{f}(\tau)$ with $\Gamma=1,i\gamma_{5},i\vec{\gamma},i\vec{\gamma}\gamma_{5}$, respectively.
They are defined at the position of the heavy quark, i.e., $\vec{x}=\vec{0}$.
When the mixing is favored energetically in the system, their expectation values
should take nonzero values.

The auxiliary fields are introduced through the following identity relation,
\begin{eqnarray}
&&
\hspace{-1em}
   \int
   {\cal D}\Phi {\cal D}\Phi^{\dag} \, {\cal D}\Phi_{5} {\cal D}\Phi_{5}^{\dag} \,
   {\cal D}\vec{\Phi} {\cal D}\vec{\Phi}^{\dag} \, {\cal D}\vec{\Phi}_{5} {\cal D}\vec{\Phi}_{5}^{\dag}
   \nonumber \\ && \hspace{-1em} \times 
   \exp
   \biggl(
       - \frac{1}{G}
         \int \drm^{4} x
         \Bigl( \Phi^{\dag}\Phi + \Phi_{5}^{\dag}\Phi_{5} + \vec{\Phi}^{\dag}\vec{\Phi}
          + \vec{\Phi}_{5}^{\dag}\vec{\Phi}_{5} \Bigr)
         \delta(\vec{x})
   \biggr)
   \nonumber \\ 
&=& {\cal N},
\label{eq:auxiliary_fields}
\end{eqnarray}
with a constant ${\cal N}$,
where
the $\delta$-function, $\delta(\vec{x})$, is inserted because $\Phi(x)$, $\Phi_{5}(x)$, $\vec{\Phi}(x)$, and $\vec{\Phi}_{5}(x)$ are essentially defined at $\vec{x}=\vec{0}$.
Multiplying the equation~\eqref{eq:auxiliary_fields}
to
the right-hand side of Eq.~\eqref{eq:partition_function_0} and shifting the auxiliary fields as
   $\Phi(x) \rightarrow \Phi(x) + G \, \bar{\psi}(x) \hat{f}(\tau)$,
   $\Phi_{5}(x) \rightarrow \Phi_{5}(x) + G \, \bar{\psi}(x) i\gamma_{5} \hat{f}(\tau)$,
   $\vec{\Phi}(x) \rightarrow \vec{\Phi}(x) + G \, \bar{\psi}(x) i\vec{\gamma} \hat{f}(\tau)$, and
   $\vec{\Phi}_{5}(x) \rightarrow \vec{\Phi}_{5}(x) + G \, \bar{\psi}(x) i\vec{\gamma}\gamma_{5} \hat{f}(\tau)$,
we rewrite the partition function $Z$ as
\begin{eqnarray}
Z
&=&
   \int
   {\cal D}\Phi {\cal D}\Phi^{\dag} \, {\cal D}\Phi_{5} {\cal D}\Phi_{5}^{\dag} \,
   {\cal D}\vec{\Phi} {\cal D}\vec{\Phi}^{\dag} \, {\cal D}\vec{\Phi}_{5} {\cal D}\vec{\Phi}_{5}^{\dag} {\cal D}\lambda
   \nonumber \\ && \times 
   \exp
   \Biggl(
         \Tr \, \ln
         S^{-1}
       - \frac{1}{G} \int \drm\tau \,
         \bigl( \Phi^{\dag}\Phi + \Phi_{5}^{\dag}\Phi_{5}
         \nonumber \\ && \hspace{3em} 
         + \vec{\Phi}^{\dag}\vec{\Phi} + \vec{\Phi}_{5}^{\dag}\vec{\Phi}_{5} \bigr)
      + \int \drm\tau \, \lambda
   \Biggr),
\label{eq:partition_function_2}
\end{eqnarray}
with the inverse of the propagator
\begin{eqnarray}
   S(x)^{-1}
=
   \left(
   \begin{array}{cc}
     \gamma\partial + m - \mu\gamma_{4}
     & 
     \bar{\Delta}(x)
     \frac{1+\gamma_{4}}{2} \delta(\vec{x}) \\
     \frac{1+\gamma_{4}}{2}
     \Delta(x)
     \delta(\vec{x})
     & \frac{1+\gamma_{4}}{2} \bigl( \partial_{\tau} + \lambda \bigr) \delta(\vec{x})
   \end{array}
   \right),
\label{eq:propagator_inverse}
\end{eqnarray}
and the gap function
\begin{eqnarray}
   \Delta(x)
&=&
   \Phi(x) + \Phi_{5}(x) i\gamma_{5} + \vec{\Phi}(x) i\vec{\gamma} + \vec{\Phi}_{5}(x) i\vec{\gamma}\gamma_{5},
\label{eq:Delta_barDelta_def}
\end{eqnarray}
and $\bar{\Delta}(x)=\gamma_{4}\Delta^{\dag}(x)\gamma_{4}$.
$\partial_{\tau}$ is a derivative with respect to $\tau$.
We notice that $S^{-1}(x)$ is the $8\times8$ dimensional matrix, in which
the diagonal $4\times4$ dimensional submatrices
 represent the propagators of a light quark and a heavy quark,
and the off-diagonal $4\times4$ dimensional submatrices
 represent the mixing between a light quark and a heavy quark.

Assuming the mean-field approximation for the composite fields $\Phi$, $\Phi_{5}$, $\vec{\Phi}$, and $\vec{\Phi}_{5}$ in Eq.~\eqref{eq:partition_function_2}, we obtain the free energy $F$ at zero temperature\footnote{We note that $\Phi$, $\Phi_{5}$, $\vec{\Phi}$, and $\vec{\Phi}_{5}$ are the fields dependent only on the Euclidean time. Here ``mean-field" means to take a constant value on the axis of the Euclidean time $\tau$ at $\vec{x}=\vec{0}$, not in the whole three-dimensional space~\cite{Hewson}.},
\begin{eqnarray}
   F
&=&
 - \Tr \, \ln
   S(x)^{-1}
+ \frac{1}{G} \int \drm\tau \,
   \bigl( \Phi^{\dag}\Phi + \Phi_{5}^{\dag}\Phi_{5}
   \nonumber \\ && \hspace{0em} 
   + \vec{\Phi}^{\dag}\vec{\Phi} + \vec{\Phi}_{5}^{\dag}\vec{\Phi}_{5} \bigr)
 - \int \drm\tau \, \lambda,
\label{eq:free_energy}
\end{eqnarray}
where
 $\Phi$, $\Phi_{5}$, $\vec{\Phi}$, $\vec{\Phi}_{5}$, and $\lambda$
are regarded
  as the classical fields which are determined by the stationary condition of the free energy.

\subsection{QCD Kondo condensates}
\label{sec:QCD_Kondo_condensate}

The gap function $\Delta$ in Eq.~\eqref{eq:Delta_barDelta_def}
is related to
the bifermion form, $\hat{f}\Gamma\psi$
with an appropriate
matrix
$\Gamma$,
in correspondence to
 the scalar ($\Phi$), the pseudoscalar ($\Phi_{5}$), the vector ($\vec{\Phi}$), and the axialvector ($\vec{\Phi}_{5}$)
boson fields.
In the following, we consider the Kondo condensate by assuming that some of $\Phi$, $\Phi_{5}$, $\vec{\Phi}$, and $\vec{\Phi}_{5}$
take non-zero values
 in the mean-field approximation,
and investigate several different types of the Kondo condensate.

Firstly,
we consider conventional configurations characterized by only
either of the
scalar condensate $\Phi$ or the pseudoscalar condensate $\Phi_{5}$, i.e.,
\begin{eqnarray}
N_{+}&:& \ \Phi\neq0\ \ {\rm while}  \ \vec{\Phi} = \vec{0}, \ \Phi_{5} = 0, \ \vec{\Phi}_{5} = \vec{0}, \label{eq:NpmDef_1} \\
N_{-} &:& \ \Phi_{5}\neq0\ \ {\rm while} \ \Phi = 0, \ \vec{\Phi} = \vec{0}, \ \vec{\Phi}_{5} = \vec{0}. \label{eq:NpmDef_2}
\end{eqnarray}
In the present paper, we call them $N_+$ and $N_-$ condensates, where the subscripts ($\pm$) represent the corresponding parity eigenvalues.
The $N_{\pm}$ condensates
give
one of the
simplest configurations, where equally both the particle and antiparticle components of the light quark correlate with
the particle component of the heavy one.\footnote{The $N_{+}$ condensate was discussed for the heavy quarks distributed uniformly in three-dimensional space, which is regarded as the Kondo lattice in the small limit of the lattice spacing~\cite{Kanazawa:2020xje,Kanazawa:2016ihl}.}

Secondly, we consider the so-called the hedgehog configurations of the Kondo condensate.
From the previous studies~\cite{Yasui:2016svc,Yasui:2017izi,Yasui:2016yet,Suzuki:2017gde}, we have learned that 
the hedgehog configuration 
 can be realized
  at sufficiently large $\mu$.
Following this fact, in the present study we invent similar ansatzes as ground states. Taking into account the condensate structures with different parities,
the hedgehog
configurations are expressed in momentum space by
\begin{eqnarray}
\hspace{-1em}
P_{+} &:&
\vec{\Phi} = -\frac{\vec{p}}{E_{\vec{p}}+m} \Phi\neq \vec{0} \ {\rm while} \ \Phi_{5}=0, \ \vec{\Phi}_{5} = \vec{0}, \label{eq:PpmDef_1} \\
\hspace{-1em}
P_{-} &:&
\vec{\Phi}_{5} = -\frac{i\vec{p}}{E_{\vec{p}}-m} \Phi_{5}\neq \vec{0} \ {\rm while} \ \Phi=0, \ \vec{\Phi} = \vec{0}, \label{eq:PpmDef_2}
\end{eqnarray}
with the three-dimensional momentum $\vec{p}$ and the energy $E_{\vec{p}}=\sqrt{\vec{p}^{2}+m^{2}}$ ,
 where the subscripts ($\pm$) represent the corresponding parity eigenvalues.
Here, we have named the configurations $P_\pm$ condensates, because they pick up only the {\it particle component} of the light quark as discussed shortly.
The $P_{+}$ (or $P_{-}$) condensate exhibits the {\it hedgehog} type in momentum space, because the direction of $\vec{\Phi}$ (or $\vec{\Phi}_{5}$) coincides with the direction of $\vec{p}$.

Using the Eqs.~\eqref{eq:PpmDef_1} and \eqref{eq:PpmDef_2}, we find that the gap function
$\Delta$ in Eq.~\eqref{eq:Delta_barDelta_def} can be rewritten in the momentum space as
\begin{eqnarray}
P_{+} &:& \hat{f}^{\dag}\Delta\psi = \frac{2E_{\vec{p}}}{E_{\vec{p}}+m} \hat{f}^{\dag}\Phi\Lambda_{\rm P}(\vec{p})\psi, \label{eq:DeltaPpm_1} \\
P_{-} &:& \hat{f}^{\dag}\Delta\psi = \frac{-2E_{\vec{p}}}{E_{\vec{p}}-m} \hat{f}^{\dag}i\gamma_{5}\Phi_{5} \Lambda_{\rm P}(\vec{p})\psi. \label{eq:DeltaPpm_2}
\end{eqnarray}
In the above equations, $\Delta$ needs to be sandwiched between the heavy quark and the light quark as $\hat{f}^{\dag}\Delta\psi$,
 because Eqs.~\eqref{eq:DeltaPpm_1} and \eqref{eq:DeltaPpm_2} hold only within the projected space, as explained shortly.
Here we defined\footnote{In the literature, understanding the hedgehog configuration as the particle-projection for a light quark was firstly pointed out 
 in Ref.~\cite{Suenaga:2020oeu}.}
\begin{eqnarray}
\Lambda_{\rm P}(\vec{p}) = \frac{-i\vec{p}\!\cdot\!\vec{\gamma}+E_{\vec{p}}\gamma_{4}+m}{2E_{\vec{p}}}, \label{eq:projection}
\end{eqnarray}
which is
 the projection operator picking up the particle component of the light quark as
indicated by
$\hat{f}^{\dag}\Delta\psi$.\footnote{In Eq.~\eqref{eq:projection}, ``$-i$'' in front of $\vec{p}\!\cdot\!\vec{\gamma}$ is put since $\gamma_k$ is defined as a hermitian matrix in the present Euclidean notation as seen from Eq.~\eqref{GammaDef}.}
For
obtaining Eqs.~\eqref{eq:DeltaPpm_1} and \eqref{eq:DeltaPpm_2}, we have implicitly used a fact that the particle-projection operator for the heavy quark: $(1+\gamma_4)/2$ is present at the left of $\Delta$ inside $\hat{f}^{\dag}\Delta\psi$.
In the followings, we introduce the assumption that $\Phi$ and $\Phi_{5}$ are factorized as $\Phi=\sqrt{(E_{\vec{p}}+m)/E_{\vec{p}}}\tilde{\Phi}$ and $\Phi_{5}=\sqrt{(E_{\vec{p}}-m)/E_{\vec{p}}}\tilde{\Phi}_{5}$ with $\tilde{\Phi}$ and $\tilde{\Phi}_{5}$ being constant values, respectively, in the momentum space.\footnote{This factorization was adopted in the case of the uniformly distributed impurities in Ref.~\cite{Suzuki:2017gde}, where the same form for the factorization was used.}

Similarly to the $P_\pm$ condensates, one can also consider the antiparticle-projected ($AP_{\pm}$) condensates picking up the {\it antiparticle component} of the light quark, which is shown in Appendix~\ref{sec:antiparticle-projected_condensates_def}. Those configurations are, however, found to be always disfavored for realistic parameter sets.

We remember that the $N_{\pm}$ condensates contain both light particles and light antiparticles on an equal footing, while the $P_{\pm}$ condensates contain light particles only. Then, it may be expected that the latter is more favored than the former as far as only the Fermi surface is concerned. However, the situation can be different in the presence of chiral symmetry breaking, as discussed below.\footnote{In the present study, for simplicity, we do not consider the possible coexistence or the competition of $N_{\pm}$ and $P_{\pm}$. Such phenomena will be investigated by minimizing the free energy with both condensates.}

\subsection{Impurity energy}
\label{sec:effective_potential}

We discuss the stability of the Kondo condensates, $N_{\pm}$ and $P_{\pm}$ condensates, introduced in Sec.~\ref{sec:QCD_Kondo_condensate}.
As a quantity measuring the stability of the Kondo condensate, we define a useful quantity,
\begin{equation}
\delta F \equiv F + \Tr \, \ln S_0(x)^{-1},
\label{eq:impurity_energy}
\end{equation}
by subtracting the free energy of noninteracting light quarks, $-\Tr \, \ln S_{0}(x)^{-1}$, from the free energy in Eq.~\eqref{eq:free_energy}. 
Here $S_{0}(x)^{-1}$ is defined by Eq.~\eqref{eq:propagator_free} in Appendix~\ref{sec:calculation_free_energy}.

We call $\delta F$ {\it impurity energy}, because
$\delta F$ measures an energy decrease due to the formation of the QCD Kondo condensate.
For example, a negative (positive) $\delta F$ means an energy decrease (increase).
In the following, we denote $\delta F$
for
 the
$N_{\pm}$, $P_{\pm}$, and $AP_{\pm}$
condensates by
$\delta F_{N_{\pm}}$, $\delta F_{P_{\pm}}$, and $\delta F_{AP_{\pm}}$,
respectively.

The impurity energies in the $N_{\pm}$ condensates are
\begin{eqnarray}
   \delta F_{N_{\pm}}
&=&
   \frac{2N_{c}}{\pi}
   \int_{m}^{\mu} \drm \omega \,
   \atan \!
   \left(
   \frac
   {
               \dfrac{1}{4\pi} (\omega+m) \sqrt{\omega^{2}-m^{2}}
               \phi 
   }
   {
               \omega-\mu-\lambda
             - \dfrac{{\cal I}_{N_{\pm}}(\omega)}{4\pi^{2}}
               \phi 
   }
   \right)
   \nonumber \\ && \times 
   \theta(\mu-m)
   \nonumber \\ && \hspace{0em} 
 - \frac{2N_{c}}{\pi}
   \int_{-\Lambda_{m}}^{-m} \drm \omega \,
   \atan \!
   \left(
   \frac
   {
               \dfrac{1}{4\pi} (\omega-m) \sqrt{\omega^{2}-m^{2}}
               \phi 
   }
   {
               \omega-\mu-\lambda
             - \dfrac{{\cal I}_{N_{\pm}}(\omega)}{4\pi^{2}}
               \phi 
   }
   \right)
   \nonumber \\ && \hspace{0em} 
+ \frac{1}{G}
   \phi 
+ 2N_{c} \lambda \theta(-\lambda)
 - \lambda,
\label{eq:eff_pot_Lambda_0}
\end{eqnarray}
with 
$\phi=\Phi^{\dag}\Phi$ for the $N_{+}$ condensate and $\phi=\Phi_{5}^{\dag}\Phi_{5}$ for the $N_{-}$ condensate.
$\theta(\cdot)$ is a step function.
The energy integral is performed both for positive energy $\omega \in [m,\mu]$ and for negative energy $\omega \in [-\Lambda_{m},-m]$, because the $N_{\pm}$ condensates include both particle-component and antiparticle-component of the light quark.
Here we define $\Lambda_{m}=\sqrt{\Lambda^{2}+m^{2}}$ with the momentum cutoff $\Lambda$ for negative-energy states in the Dirac sea of the light quarks.
The 
functions
 ${\cal I}_{N_{+}}(\omega)$ and ${\cal I}_{N_{-}}(\omega)$ are
 defined
as Eqs.~\eqref{eq:IN+_def} and \eqref{eq:IN-_def}, respectively,
in Appendix~\ref{sec:calculation_free_energy}.
Notice that
 the $N_{+}$ and $N_{-}$ condensates have different impurity energies due to ${\cal I}_{N_{+}}(\omega)\neq{\cal I}_{N_{-}}(\omega)$.

The impurity energies in the $P_{\pm}$ condensates are
\begin{eqnarray}
   \delta F_{P_{\pm}}
&=&
   \frac{2N_{c}}{\pi}
   \int_{m}^{\mu} \drm \omega \,
   \atan \!
   \left(
   \dfrac
   {
         \dfrac{1}{2\pi} \omega \sqrt{\omega^{2}-m^{2}}
         \tilde{\phi} 
   }
   {
         \omega-\mu-\lambda
       - \dfrac{{\cal I}_{P}(\omega)}{2\pi^{2}}
         \tilde{\phi} 
   }
   \right)
   \nonumber \\ && 
   \times \theta(\mu-m)
+ \frac{1}{G}
   \tilde{\phi} 
+ 2N_{c}
   \lambda\theta(-\lambda)
 - \lambda,
\label{eq:eff_pot_Lambda_p}
\end{eqnarray}
with 
$\tilde{\phi}=2\tilde{\Phi}^{\dag}\tilde{\Phi}$ for the $P_{+}$ condensate and $\tilde{\phi}=2\tilde{\Phi}_{5}^{\dag}\tilde{\Phi}_{5}$ for the $P_{-}$ condensate.
The details of the calculation are shown in Appendix~\ref{sec:calculation_free_energy}.
The range of the energy integral is limited to the positive energy $\omega \in [m,\mu]$ with $\mu \ge m$, because the $P_{\pm}$ condensates include only the particle-component of the light quark.
The function ${\cal I}_{P}(\omega)$ is 
defined as Eq.~\eqref{eq:IP_def}
 in Appendix~\ref{sec:calculation_free_energy}.
We can confirm that $\delta F_{P_{\pm}}$ reproduces the result for the massless quark case in Ref.~\cite{Yasui:2016yet} by putting $m=0$ into Eq.~\eqref{eq:eff_pot_Lambda_p}.\footnote{In Ref.~\cite{Yasui:2016yet}, the author diagonalized the mean-field Hamiltonian directly instead of summing up the expansion series as shown in Appendix~\ref{sec:calculation_free_energy}.}

We note that, from Eq.~\eqref{eq:eff_pot_Lambda_p}, the $P_{+}$ and $P_{-}$ condensates are degenerate, not only for chirally-symmetric massless quarks ($m=0$), but also for chirally-broken massive quarks ($m \neq 0$).
The degeneracy for the latter is, however, regarded to be accidental, because the chiral symmetry breaking does not necessarily lead to the nondegeneracy between the positive-parity state and the negative-parity state.\footnote{It is true that the nondegeneracy between eigenstates with opposite parity is caused by the chiral symmetry breaking, whereas its inverse does not necessarily hold in general.}

Similarly, the impurity energies for the $AP_{\pm}$ condensates, $\delta F_{AP_{\pm}}$, are shown in Appendix~\ref{sec:antiparticle-projected_condensates_energy}.
However, we do not consider the $AP_{\pm}$ condensates seriously, because they lead to only trivial states with vanishing condensates when realistic parameter sets are used.\footnote{See Sec.~\ref{sec:numerical_results} for the possible realization of the $AP_{\pm}$ condensates when an unrealistic parameter set is used.}

\section{Results}
\label{sec:results}

\subsection{Impurity energy}
\label{sec:numerical_results}

\begin{figure}[t]
\begin{center}
\vspace{0em}
\includegraphics[scale=0.28]{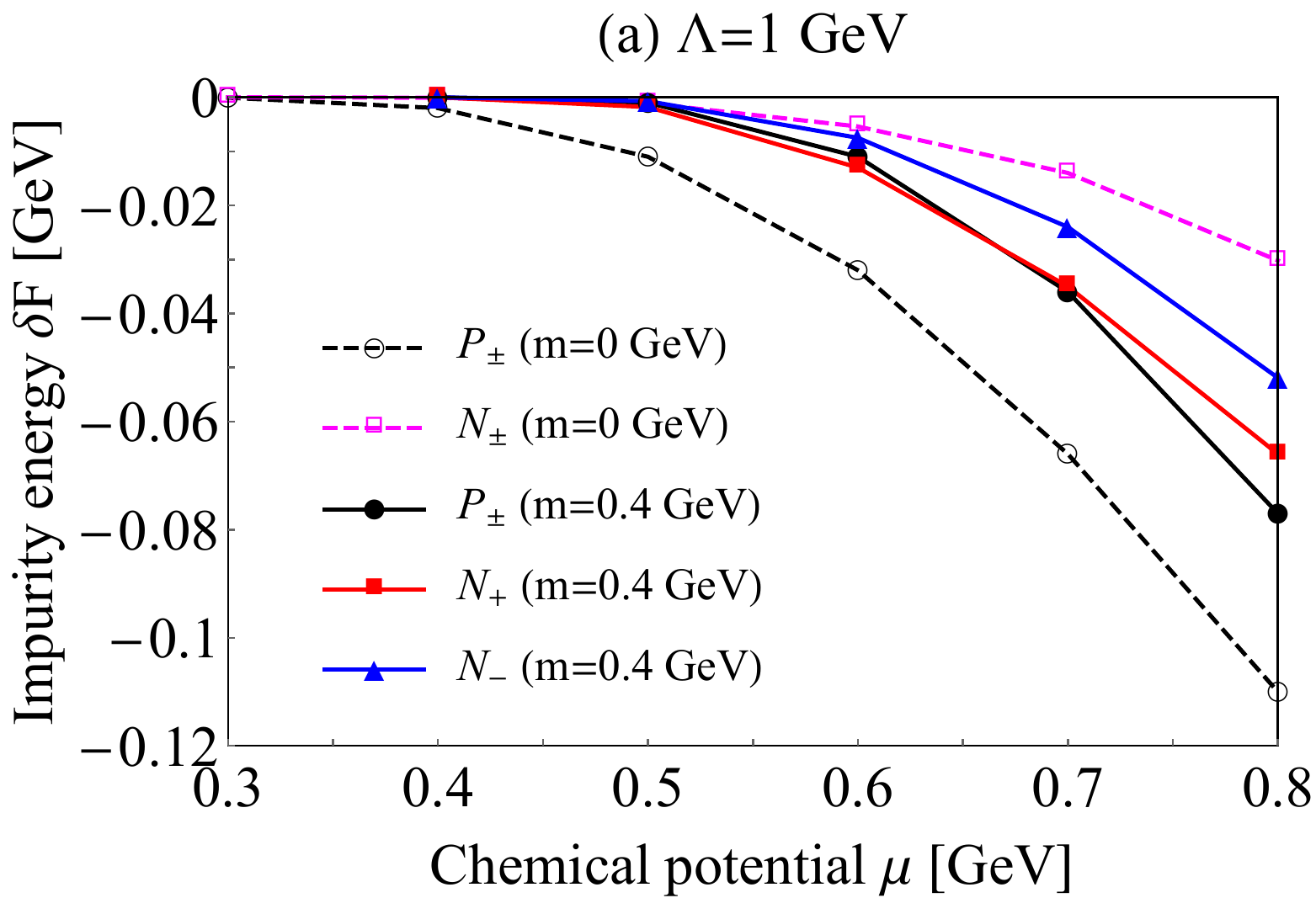}
\\
\vspace{1em}
\includegraphics[scale=0.28]{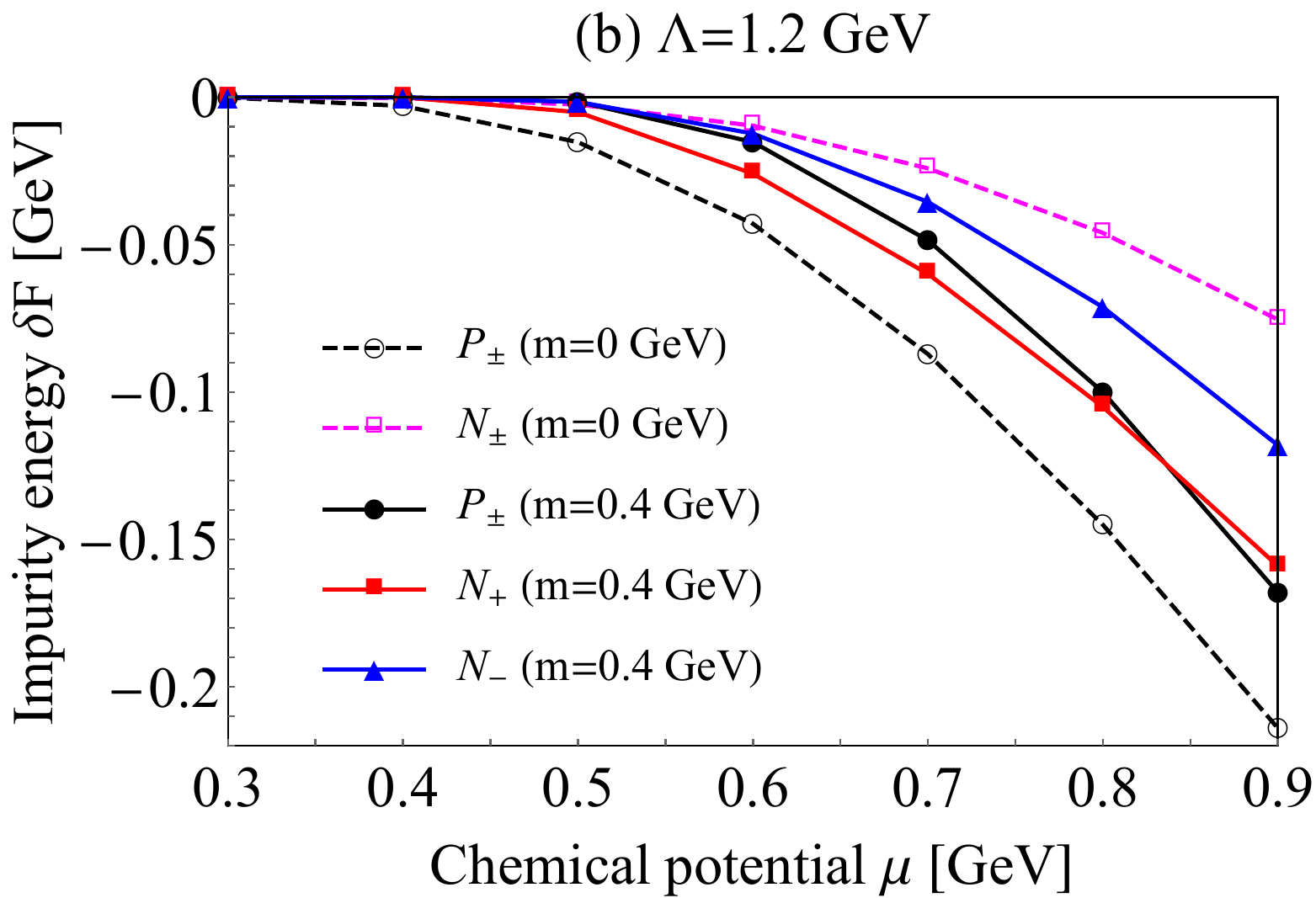}
\caption{The impurity energies $\delta F=\delta F_{N_{\pm}}, \delta F_{P_{\pm}}$ are shown
 as functions of the chemical potential $\mu$. The cutoff parameters are (a) $\Lambda=1$ GeV and (b) $\Lambda=1.2$ GeV.  The masses of a light quark are $m=0$ GeV (dashed lines) and $m=0.4$ GeV (solid lines).}
\label{fig:210417}
\end{center}
\end{figure}

In order
to determine the ground states
for the $N_{\pm}$ and $P_{\pm}$ condensates,
we show the numerical results of the impurity energies, $\delta F=\delta F_{N_{\pm}}$, $\delta F_{P_{\pm}}$, as functions of the chemical potential $\mu$ of the light quarks, see
Fig.~\ref{fig:210417}.
As the model parameters, we use the coupling constant 
$G \Lambda^{2}=4.3$
and
the momentum cutoffs, (a) $\Lambda=1$ GeV and (b) $\Lambda=1.2$ GeV.\footnote{
The value of $G\Lambda^{2}$ is larger than the one used in Eq.~(1) in Ref.~\cite{Yasui:2017izi} which corresponds to $G\Lambda^{2}=4.0$ in the present notation.
See also Appendix~A in this reference.
In the present calculation, we adopt a larger value for $G\Lambda^{2}$ and regard $\Lambda$ as a free parameter in order to enhance the differences for each type of condensate.
}
The two different cutoff parameters are introduced in order to investigate the contributions from the light antiparticles in the Dirac sea.
The masses of the light quark are set to be $m=0$ GeV (the massless quarks) and $m=0.4$ GeV (the massive quarks).

Let us see the result of $\Lambda=1$ GeV in Fig.~\ref{fig:210417}~(a).
In the massless case with $m=0$ GeV (dashed lines), the $P_{\pm}$ condensates are more favored than the $N_{\pm}$ condensates for any chemical potential.
Thus the $P_{\pm}$ condensate gives
the most stable state.
In the massive case with $m=0.4$ GeV (solid lines), the $N_{+}$ condensate gives the most stable state for the smaller chemical potential ($\mu \alt 0.65$ GeV), and the $P_{\pm}$ condensate becomes the most stable state for the larger chemical potential ($\mu \agt 0.65$ GeV).\footnote{Notice that the attraction in the $N_{-}$ condensate is weaker than that of the $N_{+}$ condensate for $m=0.4$ GeV. Such a difference may be caused by the breaking of chiral symmetry due to finite mass of light quarks.}
Such an inversion of the condensate type is more clearly seen for the larger cutoff momentum $\Lambda=1.2$ GeV in Fig.~\ref{fig:210417}~(b).
In this case, the $N_{+}$ condensate is the most stable state for $\mu \alt 0.8$ GeV, and the $P_{\pm}$ condensate becomes the most stable for $\mu \agt 0.8$ GeV.

Here, we explain the mechanism of the $\mu$-dependent inversion of the $P_{\pm}$ and $N_+$ condensates for the massive case. At sufficiently large $\mu$ ($\gg m$), only the particle component of the light quark is likely to correlate with the heavy impurity assisted by
 Fermi-surface effects substantially, while the antiparticle one is not. Hence, the $P_\pm$ condensates are favored. Such a selection rule to pick up only the particle component of the light quark is essentially provided by the factor of $\vec{p}\!\cdot\!\vec{\gamma}$ in the projection operator~\eqref{eq:projection}. 

When the chemical potential is close to the light-quark mass, $\mu \agt m$, the Fermi momentum $p_F\equiv\sqrt{\mu^2-m^2}$ gets small, and contributions from $\vec{p}\!\cdot\!\vec{\gamma}$ in the $P_{\pm}$ condensates are suppressed. Accordingly, the configuration of the $P_{\pm}$ condensates becomes similar to the $N_+$ one. Meanwhile, from the second term in the impurity energy~\eqref{eq:eff_pot_Lambda_0}, one can see that the $N_+$ condensate incorporates an additional attraction from the antiparticle component of the light quark that the $P_\pm$ condensates do not. As a result, the $N_+$ condensate is more favored than the $P_\pm$ ones in such a low-density region. For $\mu<m$, there is no energy decrease leading to the $P_\pm$ condensates, and obviously the $N_+$ one is favored.

It should be noted that the attraction from the antiparticle component for the $N_+$ condensate depends on the value of the momentum cutoff $\Lambda$. In fact, we can see that, comparing the panels (a) and (b) in Fig.~\ref{fig:210417}, the energy decrease by the condensate for the larger momentum cutoff ($\Lambda=1.2$ GeV) is larger than that of the smaller momentum cutoff ($\Lambda=1$ GeV). This will be also discussed in terms of the Kondo resonance in Sec.~\ref{sec:QCD_Kondo_resonance}.

We have shown that the QCD Kondo condensate is realized in finite density at nonzero chemical potential ($\mu>0$) within our choices of parameters.
Here, we comment on the question whether the QCD Kondo condensate could occur in vacuum ($\mu=0$),
as this problem was raised in Ref.~\cite{Kanazawa:2016ihl}.
In the present study, for $\mu=0$, we
confirmed
that the $N_{\pm}$ and $AP_{\pm}$ condensates
occur for an extremely large value of the momentum cutoff $\Lambda$, while the $P_{\pm}$ condensate
vanishes due to the absence of the antiparticle component.
However, we regard that this choice of parameter would be unrealistic in the QCD vacuum, because there is no observation of the QCD Kondo effect in vacuum.

\subsection{Interpretation: QCD Kondo resonances}
\label{sec:QCD_Kondo_resonance}

The QCD Kondo condensate
can be interpreted in terms of the resonance which is dynamically formed by the mixing between the light and heavy quarks near the Fermi surface.\footnote{See~Ref.~\cite{Yasui:2016yet} for the QCD Kondo resonance in the $P_{+}$ condensate for massless light quarks for an isolated impurity as discussed in the present study.}
The idea of the resonance picture has been discussed as the Kondo resonance (or cloud) in metals, as such studies were originally
initiated by Abrikosov~\cite{Abrikosov1965} and Suhl~\cite{Suhl1965} and later developed in the mean-field theory~\cite{Coleman1984,Read1983a,Read1983b} (see, e.g., Ref.~\cite{Hewson} and references therein).

We show the resonance picture in the present study. For this purpose, we approximate the impurity energies  \eqref{eq:eff_pot_Lambda_0} and \eqref{eq:eff_pot_Lambda_p} at a large chemical potential.
As an approximation, we replace the denominator of the arctangent functions in Eqs.~\eqref{eq:eff_pot_Lambda_0} and \eqref{eq:eff_pot_Lambda_p} with $\omega-\mu-\lambda$ by neglecting ${\cal I}_{i}(\omega)$
($i=N_{\pm}$, $P$),
and we replace $\omega$ in the numerator with $\mu$ because the energy around the Fermi surface $(\omega \approx \mu)$ gives the most dominant contribution to the energy integrals.

For the $N_{\pm}$ condensates, the impurity energy~\eqref{eq:eff_pot_Lambda_0}
 is approximated as
\begin{eqnarray}
   \delta F_{N_{\pm}}
&\approx&
   2N_{c}
   \int_{m}^{\mu} \drm \omega \,
   (\omega-\mu)
   \rho_{N}(\omega)
   \theta(\mu-m)
   \nonumber \\ && 
+ 2N_{c}
   \int_{-\Lambda_{m}}^{-m} \drm \omega \,
   (\omega-\mu)
   \rho_{N}(\omega)
+ \frac{1}{G}
   \phi 
   \nonumber \\ && 
+ 2N_{c} \lambda \theta(-\lambda)
 - \lambda,
\label{eq:eff_pot_Lambda_0_2}
\end{eqnarray}
where the spectral function $\rho_{N}(\omega)$ is defined as
\begin{eqnarray}
   \rho_{N}(\omega)
&=&
   \frac{1}{\pi}
   \dfrac{\Delta_{N}}{(\omega-\mu-\lambda)^{2}+\Delta_{N}^{2}},
\label{eq:spectral_function_N} 
\end{eqnarray}
with $\Delta_{N} =
   \pi
   (\mu+m) \sqrt{\mu^{2}-m^{2}} \,
   \phi 
   /(4\pi^{2})
$.
This result indicates the Lorentzian-type resonance with the energy position $\mu+\lambda$ and the width $\Delta_{N}$.
Thus, we understand that the QCD Kondo condensate gives the width of the QCD Kondo resonance.
We notice that the impurity energies of the $N_{+}$ and $N_{-}$ condensates
coincide with each other
accidentally in the present approximation, because ${\cal I}_{N_{\pm}}(\omega)$ in Eq.~\eqref{eq:eff_pot_Lambda_0} have been neglected.
Thus this degeneracy should not be taken seriously.

For the $P_{\pm}$ condensate, the impurity energy~\eqref{eq:eff_pot_Lambda_p} is approximated as
\begin{eqnarray}
   \delta F_{P_{\pm}}
&\approx&
   2N_{c}
   \int_{m}^{\mu} \drm \omega \,
   (\omega-\mu)
   \rho_{P}(\omega)
   \theta(\mu-m)
+ \frac{1}{G}
   \tilde{\phi} 
   \nonumber \\ && 
+ 2N_{c}
   \lambda\theta(-\lambda)
 - \lambda,
\label{eq:eff_pot_Lambda_p_2}
\end{eqnarray}
where
 the spectral function $\rho_{P}(\omega)$ is defined as
\begin{eqnarray}
   \rho_{P}(\omega)
&=&
   \frac{1}{\pi}
   \dfrac{\Delta_{P}}{(\omega-\mu-\lambda)^{2}+\Delta_{P}^{2}},
\label{eq:spectral_function_P}
\end{eqnarray}
with
$\Delta_{P} =
   \pi
   \mu \sqrt{\mu^{2}-m^{2}} \,
   \tilde{\phi} 
   /(2\pi^{2})
$.
Thus
we obtain the resonance state with the energy position $\mu+\lambda$ and the width $\Delta_{P}$.
We see again that the QCD Kondo condensate gives the width of the QCD Kondo resonance.
For the $P_{\pm}$ condensates, we can prove that there is always a nonzero value of the QCD Kondo condensate for any small (attractive) coupling, as shown in Appendix~\ref{sec:simple_solution}.

In summary, the QCD Kondo condensate is regarded as the resonance state with the energy position $\lambda$ above the Fermi surface and the width $\Delta_{i}$
($i=N$, $P$).
We call it {\it QCD Kondo resonance} (or {\it QCD Kondo cloud}).
This resonance is stabilized by the energy decrease through the mixing between a light quark and a heavy quark.
The QCD Kondo resonance is analogous to the Kondo resonance
 in condensed matter systems, which is caused by the mixing between an itinerant (conducting) electron and a localized electron~\cite{Gruner1974,Gubernatis1987,Affleck1996,Affleck2010}.\footnote{Recently the experimental observation of the Kondo cloud was reported~\cite{V.Borzenets2020}.}

\subsection{$N_{+}$ condensate v.s. $P_{\pm}$ condensate}
\label{sec:condensate_comparison}

In Sec.~\ref{sec:numerical_results}, we discussed based on the numerical computations the inversion of the $N_{+}$ and $P_{\pm}$ condensates for large chemical potentials $\mu \agt m$ (see Fig.~\ref{fig:210417}).
In this subsection, we show that this inversion can be understood also at a qualitative level in terms of the widths of the QCD Kondo resonance.

Since the energy position $\omega=\mu+\lambda$ of the Kondo resonance always lies above the integration range $m<\omega<\mu$ for the particle-component contributions as seen from Eq.~\eqref{eq:eff_pot_Lambda_0_2} [or Eq.~\eqref{eq:spectral_function_N}], one can reasonably infer that the larger value of the 
resonance width ($\Delta_{N}$ or $\Delta_{P}$) leads to the greater energy decrease.
Since the width is proportional to the magnitude of the condensate,
we expect that a more favored (stable) condensate can be realized for more enhanced resonance width.

In order to check this property with respect to the $N_{+}$ and $P_{\pm}$ condensates in an intuitive manner, 
we focus on the ratio of the widths $\Delta_{N}$ and $\Delta_{P}$ in Eqs.~\eqref{eq:spectral_function_N} and \eqref{eq:spectral_function_P},
\begin{equation}
\frac{\Delta_{P}}{\Delta_{N}} = \frac{ 2}{1+m/\mu}.
\label{eq:ratio_NP}
\end{equation}
From this ratio, we can achieve the following reinterpretation of the ground-state configurations:
\begin{enumerate}
\item When the chemical potential $\mu$ is sufficiently large ($\mu \gg m$) at high density,
the
ratio
\eqref{eq:ratio_NP}
approaches two, i.e., $\Delta_{P}/\Delta_{N} \approx 2$.
Thus, the resonance in the $P_{\pm}$ condensates has a width almost twice as large as that of the $N_{+}$ condensate, and hence the $P_{\pm}$ condensates are more stable than the $N_{+}$ condensate.
\item In contrast, when $\mu$ is close to the light-quark mass $m$ threshold ($\mu \agt m$) at low density,
the ratio~\eqref{eq:ratio_NP} approaches one, i.e., $\Delta_{P}/\Delta_{N} \agt 1$.
Then, one might think that the $N_{+}$ condensate and the $P_{\pm}$ condensate are almost same.
Nevertheless,
the additional energy decrease is provided for the $N_{+}$ condensate by the attraction from the light antiparticle component in the second term of Eq.~\eqref{eq:eff_pot_Lambda_0_2}, and it makes the $N_{+}$ condensate more stable than the $P_{\pm}$ condensates.
\end{enumerate}
From the above consideration, we find that the inversion of the $N_+$ and $P_\pm$ condensates is understood qualitatively through magnitude of the width of QCD Kondo resonances.

\section{Conclusion and outlooks}

We have discussed the QCD Kondo effect on the heavy quark as a single impurity particle embedded in the quark matter of massive light quarks with chiral symmetry breaking.
In order to understand the ground state nonperturbatively,
we have introduced the QCD Kondo condensate that measures the strength of mixing between a light quark and a heavy quark.
We have considered the scalar, pseudoscalar, vector, and axialvector channels in the interaction between the light quark and the heavy quark, and have investigated different types of condensates, i.e., the $N_{\pm}$, $P_{\pm}$, and $AP_{\pm}$ condensates.
For massive light quarks,
we have shown that the most favored condensate is the $N_{+}$ condensate at small chemical potential, while the $P_{\pm}$ condensate is favored at large chemical potential.
The $AP_{\pm}$ condensate is disfavored at any chemical potential for realistic parameter sets.

In the present study, we have not specified the origin of the light quark mass: it can be the current mass in the QCD Lagrangian or the dynamical mass generated by the spontaneous chiral-symmetry breaking.
Among the 
quarks with light flavors (up, down, and strangeness), the lightest quarks, i.e., the up quarks, would be most favored to make the QCD Kondo condensate when the chemical potentials are near the masses of the light quark.
This is clearly suggested from the result that the zero mass is most favored to form the QCD Kondo condensate, as shown in Fig.~\ref{fig:210417}.
As for the dynamical masses, we can regard in the present study that
the chiral-symmetry broken state
 can coexist with the QCD Kondo condensate ($N_{+}$ or $P_{\pm}$).
This is essentially the same as the case that the heavy quarks are distributed uniformly in three-dimensional space, as studied for the $N_{+}$ condensate~\cite{Kanazawa:2020xje} and for the $P_{+}$ condensate~\cite{Suzuki:2017gde}.

We comment on several possible extensions for future studies.
Throughout the present paper, we have assumed that the light-quark mass is a constant value independent of the spatial position.
More generally, however, the dynamical mass, characterized by the chiral condensate, can depend on the position, i.e., near the heavy quark or in the bulk space away from the heavy quark.
Since our present result shows that the QCD Kondo condensate favors the smaller mass of the light quark, it may lead to the partial restoration of the chiral symmetry around the heavy quark.
In order to clarify this problem, we need to evaluate the position dependence of the chiral condensate in a self-consistent way.
Furthermore, we have neglected the interaction between two light quarks relevant to the color superconductivity characterized by diquark condensates.
It will be interesting to investigate the position dependence of diquark condensates around a single heavy quark.

There are still many open problems regarding the QCD Kondo effect:
\begin{itemize}

\item In the QCD phase diagram at finite density, the Kondo condensate can compete with the chiral and/or diquark condensates.
In the low-density phase, the competition with the (uniformly distributed) chiral condensate can be studied in the mean-field theory~\cite{Suzuki:2017gde,Ishikawa:2021bey,Hattori:2022lnh} and in the random-matrix theory~\cite{Kanazawa:2020xje}.
At a high density, the competition with the color superconductivity was studied in Ref.~\cite{Kanazawa:2016ihl}.\footnote{See, e.g., Ref.~\cite{RevModPhys.78.373} for a review on the competition of the superconductors with the impurities in condensed matter systems.
In a color superconducting phase, gapped light quarks coexist with ungapped quarks and play the role of impurities, which can lead to an emergent QCD Kondo effect~\cite{Hattori:2019zig}.}
\item The transport coefficients under the QCD Kondo effect, such as the electric resistivity and the shear viscosity, were calculated in Ref.~\cite{Yasui:2017bey}.
Also, the QCD Kondo excitons~\cite{Yasui:2017izi,Suenaga:2019car} can be excited from the ground state and additionally contribute to transport coefficients.\footnote{In Ref.~\cite{Suenaga:2019car}, the random-phase-approximation equation for the QCD Kondo excitons, which was partly studied in Ref.~\cite{Yasui:2017izi}, was studied fully by considering all the relevant channels.}
\item In relativistic heavy-ion collisions, particles and antiparticles with heavy flavors are simultaneously generated by the pair creations, and hence the QCD Kondo effect including the effect of the heavy antiparticles should be taken into account~\cite{Araki:2020fox}.
In the mean-field theory, since one can consider condensates including heavy antiparticles, the different types of condensates from those considered in the present work may be significant.
\item When the number of light-quark flavors is $N_{f}\ge2$, the QCD Kondo effect exhibits non-Fermi liquid behaviors~\cite{Kanazawa:2016ihl} (which is a realization of the multi-channel Kondo effect~\cite{Nozieres:1980}).
For such systems, the conventional mean-field theory cannot be applied, and other approaches such as the conformal field theory~\cite{Kimura:2018vxj,Kimura:2016zyv} are needed.
\item If heavy quarks are abundantly present inside compact stars (which may be called ``charm/bottom stars" \cite{Kettner:1994zs}), the QCD Kondo effect may change the properties of stars.
A possible discussion was given in Ref.~\cite{Macias:2019vbl}.
Then, depending on the density profile of heavy quarks inside stars, both uniformly distributed impurities and single impurities will be essential.
\item The QCD Kondo effect can also be studied by numerical simulations of lattice QCD, but Monte Carlo simulations at a large chemical potential are difficult due to the sign problem.
However, although environments with a chiral chemical potential~\cite{Suenaga:2019jqu,Kanazawa:2020xje} or with a strong magnetic field~\cite{Ozaki:2015sya,Kimura:2018vxj,Hattori:2022lnh} can induce the QCD Kondo effect, they are free from the sign problem and can be examined by Monte Carlo simulations.
Thus, lattice QCD simulations are powerful tools for studying this effect.
The formulation of lattice field theory to simulate QCD with a single color impurity, as studied in the present work, will be required.
\item When the quark matter is under a magnetic field, characteristic transport phenomena can be induced, such as the modification of the chiral separation effect~\cite{Araki:2020rok,Suenaga:2020oeu} and the spin polarization of heavy impurities~\cite{Araki:2020rok,Suenaga:2021wio}.
As a result of the latter, the spin polarizations of charm ($c$) and bottom ($b$) quarks induced by the QCD Kondo effect can be mapped to the spin polarization of the $\Lambda_{c}$ and/or $\Lambda_{b}$ baryons, which are more direct observables (than non-measurable heavy quarks) in heavy-ion collisions.
\item Finally, we comment that some heavy hadrons in nuclear matter
can have non-Abelian interaction with nucleons by spin and/or isospin exchange, and hence it can lead to the spin-isospin Kondo effect in analogy to the QCD Kondo effect~\cite{Yasui:2013xr,Yasui:2016hlz,Yasui:2016ngy,Yasui:2019ogk}.\footnote{See, e.g., Ref.~\cite{Hosaka:2016ypm} for a review of the studies on heavy hadrons in nuclear matter.}
The connection of the Kondo effects between the quark matter and the nuclear matter is an interesting question.
\end{itemize}
These studies will be awaited in future.

\section*{Acknowledgements}

The authors thank Y.~Araki and K.~Hattori for fruitful discussions.
S.~Y. is supported by the Ministry of Education, Culture, Sports, Science (MEXT)-Supported Program for the Strategic Research Foundation at Private Universities Topological Science (Grant No. S1511006), by the Interdisciplinary Theoretical and Mathematical Sciences Program (iTHEMS) at RIKEN, and by the World Premier International Research Center Initiative (WPI) under MEXT, Japan.
D.~S. is supported by the RIKEN special postdoctoral researcher program.
This work is supported by the Japan Society for the Promotion of Science (JSPS) KAKENHI Grants, No.~JP17K05435 (S.~Y.), No.~23K03377  (D.~S.), No.~23H05439 (D.~S.), and No.~JP20K14476 (K.~S.).
This work was supported in part by the COREnet project of RCNP, Osaka University.

\appendix

\section{Connection to HQET Lagrangian}
\label{sec:HQET_Lagrangian}

The heavy quark can be described by the heavy-quark effective theory (HQET), i.e., an effective theory of QCD in the large mass limit of a heavy quark~\cite{Eichten:1989zv,Georgi:1990um}.\footnote{See e.g. Refs.~\cite{Neubert:1993mb,Manohar:2000dt,Casalbuoni:1996pg} for reviews and textbooks on the HQET.}
In this appendix, we show how the Lagrangian \eqref{eq:Lagrangian_fermion_LH_isolated_2} can be obtained from the framework of the HQET.
For this purpose, we start from the following Lagrangian,
\begin{eqnarray}
   {\cal L}_{\HQET}
&=&
 - \bar{\psi} \bigl( \gamma\partial + m \bigr) \psi
 - G_{c}
   \bigl( \bar{\psi} \gamma_{4} T^{a} \psi \bigr)
   \bigl( \bar{\Psi}_{v} \gamma_{4} T^{a} \Psi_{v} \bigr)
\nonumber \\ && 
 - \bar{\Psi}_{v} \partial_{\tau} \Psi_{v},
\label{eq:Lagrangian_fermion_LH}
\end{eqnarray}
in the Euclidean spacetime, where we neglect the chemical potential $\mu$ of light quarks, for brevity.
The interaction term with the coupling constant $G_{c}>0$ mimics the one-gluon exchange interaction between a light quark and a heavy quark.
The sign of $G_{c}$ is chosen to provide an attraction in the color-singlet channel.
Here, $T^{a}=\lambda^{a}/2$ is the generator of $\mathrm{SU}(N_{c})$ symmetry with the Gell-Mann matrices $\lambda^{a}$ ($a=1,2,\dots,N_{c}^{2}-1$).
$\psi(x)$ is the field of a light quark with the Dirac spinor, and $\Psi_{v}(x)$ is the effective field of a heavy quark in the HQET.
The latter is defined as $\Psi_{v}(x)=\frac{1}{2}(1+\gamma_{4})e^{iMvx}\Psi(x)$ for the original heavy-quark field $\Psi(x)$ in QCD.
Here $v_{\mu}$ is the four-velocity of the rest frame of the heavy quark, satisfying the conditions $v_{\mu}v_{\mu}=1$ (a sum over $\mu=1,2,3,4$).
We take $v_{\mu}=(\vec{0},1)$ at the rest frame where the heavy quark has no motion.

As the basic idea of the HQET, the four-momentum $p_{\mu}$ in $\Psi$ is divided into two parts: the on-mass-shell part $Mv_{\mu}$ and the off-mass-shell (residual) part $k_{\mu}$. $M$ is the mass of the heavy quark.
The magnitude of the residual four-momentum $k_{\mu}$ is assumed to be much smaller than $M$: $|k_{\mu}| \ll M$.
In the definition of the effective field $\Psi_{v}(x)$, the on-mass-shell part is extracted as a plane wave $e^{iMvx}$, and thus $\Psi_{v}(x)$ depends only on $k_{\mu}$.
In Eq.~\eqref{eq:Lagrangian_fermion_LH}, the (Euclidean) time-derivative operator $\partial_{\tau}$
 is understood to act on the residual momentum $k_{4}$ ($\mu=4$).

In the formalism of the HQET, we consider that the heavy quark locates at the position $\vec{x}=\vec{0}$.
In this situation, we can assume that the effective field $\Psi_{v}(x)$ is replaced by
\begin{eqnarray}
   \Psi_{v}(x) \rightarrow
\hat{f}(\tau)=
\left(
\begin{array}{c}
f_{\uparrow}(\tau) \\ f_{\downarrow}(\tau) \\ 0 \\ 0
\end{array}
\right)
\hspace{0.5em} \mathrm{at} \hspace{0.5em} \vec{x}=\vec{0},
\label{eq:replace_Psiv_f}
\end{eqnarray}
where $\hat{f}(\tau)$ is an effective field defined only at $\vec{x}=\vec{0}$.
We note that $\hat{f}(\tau)$ has dependence only on the (Euclidean) time $\tau$.
Because of the factor $(1+\gamma_{4})/2$ in $\Psi_{v}(x)$, $\hat{f}(\tau)$ has only the upper two spin-components, $f_{\uparrow}(\tau)$ and $f_{\downarrow}(\tau)$.
As for the color symmetry, $\hat{f}(\tau)$
 belongs to the fundamental representation.
We impose a constraint condition $\hat{f}^{\dag}(\tau)\hat{f}(\tau)=1$, i.e.,
\begin{eqnarray}
   f_{\uparrow}^{\dag}(\tau)f_{\uparrow}(\tau)+f_{\downarrow}^{\dag}(\tau)f_{\downarrow}(\tau)=1.
\label{eq:constraint_condition}
\end{eqnarray}
This condition is necessary because there should be only one heavy quark with either up or down spin and with either red, green, or blue in color.
Using Eq.~\eqref{eq:replace_Psiv_f}, we introduce the following replacement in the bilinear form of $\Psi_{v}(x)$,
\begin{eqnarray}
   \bar{\Psi}_{v}(x)\Gamma\Psi_{v}(x) \rightarrow \hat{f}^{\dag}(\tau) \Gamma \hat{f}(\tau) \delta(\vec{x}),
\label{eq:replace_Psiv_f_2}
\end{eqnarray}
for any $4\times4$ dimensional matrix $\Gamma$.
For example, $\Gamma$ indicates $\gamma_{4}\lambda^{a}$ or $\partial_{\tau}$ in Eq.~\eqref{eq:Lagrangian_fermion_LH}.
With the constraint condition~\eqref{eq:constraint_condition} and the replacement~\eqref{eq:replace_Psiv_f_2}, we change the Lagrangian \eqref{eq:Lagrangian_fermion_LH} to
\begin{eqnarray}
   {\cal L}_{\imp}
&=&
 - \bar{\psi} \bigl( \gamma\partial + m \bigr) \psi
 - G_{c}
   \bigl( \bar{\psi} \gamma_{4} T^{a} \psi \bigr)
   \bigl( \hat{f}^{\dag} \gamma_{4} T^{a} \hat{f} \, \delta(\vec{x}) \bigr)
\nonumber \\ && 
 - \hat{f}^{\dag} \partial_{\tau} \hat{f} \, \delta(\vec{x})
 - \lambda \bigl( \hat{f}^{\dag}\hat{f} - 1 \bigr) \, \delta(\vec{x}),
\label{eq:Lagrangian_fermion_LH_isolated}
\end{eqnarray}
where $\lambda$ is a Lagrange multiplier.
Then, we apply the Fierz transformation by the Fierz identities,
\begin{eqnarray}
   \sum_{\mu=1}^{4} (\gamma_{\mu})_{\alpha\beta} (\gamma_{\mu})_{\gamma\delta}
&=&
   \delta_{\alpha\delta} \delta_{\gamma\beta}
+ (i\gamma_{5})_{\alpha\delta} (i\gamma_{5})_{\gamma\beta}
\nonumber \\ && 
 - \frac{1}{2} \sum_{\mu=1}^{4} (\gamma_{\mu})_{\alpha\delta} (\gamma_{\mu})_{\gamma\beta}
\nonumber \\ && 
 - \frac{1}{2} \sum_{\mu=1}^{4} (\gamma_{\mu}\gamma_{5})_{\alpha\delta} (\gamma_{\mu}\gamma_{5})_{\gamma\beta},
\label{eq:Fierz_identity_Dirac}
\end{eqnarray}
for the Dirac matrices $\gamma_{\mu}$ ($\mu=1,2,3,4$) and
\begin{eqnarray}
   \sum_{a=1}^{N_{c}^{2}-1} (\lambda^{a})_{ij} (\lambda^{a})_{kl}
&=&
   2\dfrac{N_{c}^{2}-1}{N_{c}^{2}} \delta_{il} \delta_{kj}
\nonumber \\ && 
 - \dfrac{1}{N_{c}} \sum_{a=1}^{N_{c}^{2}-1} (\lambda^{a})_{il} (\lambda^{a})_{kj},
\label{eq:Fierz_identity_GellMann}
\end{eqnarray}
for the Gell-Mann matrices $\lambda^{a}$ ($a=1,2,\dots,N_{c}^{2}-1$), where we leave only the first term in Eq.~(\ref{eq:Fierz_identity_GellMann}) as the leading-order term in the $1/N_{c}$ expansion.
To obtain Eq.~\eqref{eq:Lagrangian_fermion_LH_isolated}, we finally multiply a minus sign due to the interchange of fermion operators.
Recovering the chemical potential $\mu$,
we obtain our
Lagrangian~\eqref{eq:Lagrangian_fermion_LH_isolated_2} with the coupling constant $G$ defined as $G=(G_{c}/4)(N_{c}^{2}-1)/N_{c}^{2}$.

\section{Antiparticle-projected ($AP_{\pm}$) condensates}
\label{sec:antiparticle-projected_condensates}

\subsection{Definition of $AP_{\pm}$ condensates}
\label{sec:antiparticle-projected_condensates_def}

In contrast to the $P_{\pm}$ condensate, in which only the particle component participates in the Kondo condensate, we consider the QCD Kondo condensate in which only the antiparticle component participates in. Such configurations are expressed by
\begin{eqnarray}
AP_{+} &:&
\vec{\Phi} = \frac{\vec{p}}{E_{\vec{p}}-m} \Phi\neq \vec{0} \ {\rm while} \ \Phi_{5}=0, \ \vec{\Phi}_{5} = \vec{0}, \label{eq:APpmDef_1} \\
AP_{-} &:&
\vec{\Phi}_{5} = \frac{i\vec{p}}{E_{\vec{p}}+m} \Phi_{5}\neq \vec{0} \ {\rm while} \ \Phi=0, \ \vec{\Phi} = \vec{0}, \label{eq:APpmDef_2}
\end{eqnarray}
with $E_{\vec{p}}=\sqrt{\vec{p}^{2}+m^{2}}$
in momentum space, where the subscripts ($\pm$) represent the corresponding parity eigenvalues.
Here, we have named these configurations $AP_\pm$ condensates, because they pick up only the {\it antiparticle component} of the light quark.

Using Eqs.~\eqref{eq:APpmDef_1} and \eqref{eq:APpmDef_2}, we find that the gap function
$\Delta$ in Eq.~\eqref{eq:Delta_barDelta_def} can be written in the momentum space as
\begin{eqnarray}
AP_{+} &:& \hat{f}^{\dag}\Delta\psi = \frac{2E_{\vec{p}}}{E_{\vec{p}}-m} \hat{f}^{\dag}\Phi\Lambda_{\rm AP}(\vec{p})\psi, \label{eq:DeltaAPpm_1} \\
AP_{-} &:& \hat{f}^{\dag}\Delta\psi = \frac{-2E_{\vec{p}}}{E_{\vec{p}}+m} \hat{f}^{\dag}i\gamma_{5}\Phi_{5} \Lambda_{\rm AP}(\vec{p})\psi. \label{eq:DeltaAPpm_2}
\end{eqnarray}
In Eqs.~\eqref{eq:DeltaAPpm_1} and \eqref{eq:DeltaAPpm_2},
\begin{eqnarray}
\Lambda_{\rm AP}(\vec{p}) = \frac{i\vec{p}\!\cdot\!\vec{\gamma}+E_{\vec{p}}\gamma_{4}-m}{2E_{\vec{p}}},
\end{eqnarray}
is the projection operator picking up the antiparticle component of the light quark.
Similarly to the $P_{\pm}$ condensates, the $AP_{+}$ (or $AP_{-}$) condensate exhibits the {\it hedgehog} type in momentum space, because the direction of $\vec{\Phi}$ (or $\vec{\Phi}_{5}$) coincides with the direction of the momentum $\vec{p}$.

\subsection{Impurity energy
 in $AP_{\pm}$ condensates}
\label{sec:antiparticle-projected_condensates_energy}

The impurity energies in the $AP_{\pm}$ condensates are
\begin{eqnarray}
   \delta F_{AP_{\pm}}
&=&
 - \frac{2N_{c}}{\pi}
   \int_{-\Lambda_{m}}^{-m} \drm \omega \,
   \atan \!
   \left(
   \dfrac
   {
         \dfrac{\pi \omega \sqrt{\omega^{2}-m^{2}}}{2\pi^{2}}
         \hat{\phi}
   }
   {
         \omega-\mu-\lambda
       - \dfrac{{\cal I}_{AP}(\omega)}{2\pi^{2}}
         \hat{\phi}
   }
   \right)
   \nonumber \\ && 
+ \frac{1}{G}
   \hat{\phi}
+ 2N_{c}
   \lambda\theta(-\lambda)
 - \lambda,
\label{eq:eff_pot_Lambda_ap}
\end{eqnarray}
with $\hat{\phi}=2\tilde{\Phi}^{\dag}\tilde{\Phi}$ for the $AP_{+}$ condensate and $\hat{\phi}=2\tilde{\Phi}_{5}^{\dag}\tilde{\Phi}_{5}$ for the $AP_{-}$ condensate.
Here we consider that $\Phi$ and $\Phi_{5}$ are factorized by $\Phi=\sqrt{(E_{\vec{p}}-m)/E_{\vec{p}}}\tilde{\Phi}$ and $\Phi_{5}=\sqrt{(E_{\vec{p}}+m)/E_{\vec{p}}}\tilde{\Phi}_{5}$ with $\tilde{\Phi}$ and $\tilde{\Phi}_{5}$ being constant values, respectively, in the momentum space.
The range in the energy integral is limited to the negative energy $\omega \in [-\Lambda_{m},-m]$ with $\Lambda_{m}=\sqrt{\Lambda^{2}+m^{2}}$.
Here $\Lambda$ is the cutoff parameter in the three-dimensional momentum integral in the function ${\cal I}_{AP}(\omega)$.
The momentum cutoff is introduced because the $AP_{\pm}$ condensates include the antiparticle-component of the light quark in the Dirac sea.
The
 function ${\cal I}_{AP}(\omega)$ is
 defined as Eq.~\eqref{eq:IAP_def}
 in Appendix~\ref{sec:calculation_free_energy}.

\subsection{QCD Kondo resonance in $AP_{\pm}$ condensates}
\label{sec:antiparticle-projected_condensates_resonance}

In the $AP_{\pm}$ condensates,
 the impurity energy \eqref{eq:eff_pot_Lambda_ap} is  expressed by
\begin{eqnarray}
   \delta F_{AP_{\pm}}
&\approx&
   2N_{c}
   \int_{-\Lambda_{m}}^{-m} \drm \omega \,
   (\omega-\mu)
   \rho_{AP}(\omega)
+ \frac{1}{G}
   \hat{\phi}
   \nonumber \\ && 
+ 2N_{c}
   \lambda\theta(-\lambda)
 - \lambda,
\label{eq:eff_pot_Lambda_ap_2}
\end{eqnarray}
where the spectral function $\rho_{AP}(\omega)$ is defined as
\begin{eqnarray}
   \rho_{AP}(\omega)
&=&
   \frac{1}{\pi}
   \dfrac{\Delta_{AP}}{(\omega-\mu-\lambda)^{2}+\Delta_{AP}^{2}},
\label{eq:spectral_function_AP}
\end{eqnarray}
with $\Delta_{AP} =
   \pi
   \mu \sqrt{\mu^{2}-m^{2}} \,
   \hat{\phi}
   /(2\pi^{2})
$.
This indicates the resonance of the Lorentzian-type with the energy $\mu+\lambda$ and the width $\Delta_{AP}$.

\section{Calculation of impurity energy}
\label{sec:calculation_free_energy}

We explain the calculation of the free energy~\eqref{eq:free_energy}.
Because the treatment of the $\delta$-function ($\delta(\vec{x})$) in
the free energy
may not be commonly known in the literature, it may be valuable to present the concrete steps of the calculation.
The mean-field approximation for a single impurity was developed by several researchers in condensed matter physics~\cite{Takano:1966,Yoshimori:1970,ReadNewns1983,Eto:2001,Yanagisawa:2015conf,Yanagisawa:2015}.
First of all, we separate the propagator in Eq.~\eqref{eq:propagator_inverse} as 
\begin{eqnarray}
   S(x)^{-1}
=
   S_{0}(x)^{-1}
+ \tilde{\Delta}(x) \, \delta(\vec{x}),
\label{eq:propagator_full}
\end{eqnarray}
where the first term is the free part
\begin{eqnarray}
   S_{0}(x)^{-1}
&=&
   \left(
   \begin{array}{cc}
     \gamma\partial + m - \mu \gamma_{4} & 0 \\
     0 & \frac{1+\gamma_{4}}{2} \bigl( \partial_{\tau} + \lambda \bigr) \delta(\vec{x})
   \end{array}
   \right),
\nonumber \\
\label{eq:propagator_free}
\end{eqnarray}
and the second term is for the QCD Kondo condensate.
$\tilde{\Delta}(x)$ is defined as
\begin{eqnarray}
   \tilde{\Delta}(x)
&=&
   \left(
   \begin{array}{cc}
     0  & \bar{\Delta}(x) \frac{1+\gamma_{4}}{2} \\
     \frac{1+\gamma_{4}}{2} \Delta(x) & 0
   \end{array}
   \right),
\end{eqnarray}
with $\Delta(x)$ in Eq.~\eqref{eq:Delta_barDelta_def}.
With these setups, we expand formally the free energy~\eqref{eq:free_energy} as
\begin{eqnarray}
   F
&=&
 - \Tr \, \ln S_{0}(x)^{-1}
   \nonumber \\ && 
 - \sum_{k\ge1} \frac{(-1)^{k+1}}{k}
   \Tr
   \bigl( S_{0}(x)
   \tilde{\Delta}(x) \, \delta(\vec{x})
   \bigr)^{k}
   \nonumber \\ && \hspace{0em} 
+ \frac{1}{G} \int \drm\tau \,
   \bigl( \Phi^{\dag}\Phi + \Phi_{5}^{\dag}\Phi_{5} + \vec{\Phi}^{\dag}\vec{\Phi} + \vec{\Phi}_{5}^{\dag}\vec{\Phi}_{5} \bigr)
   \nonumber \\ && 
 - \int \drm\tau \, \lambda.
\label{eq:eff_pot_Kondo_single_GL}
\end{eqnarray}
The first term in 
the right-hand side
 is relevant only to the light quarks and it is irrelevant to the QCD Kondo condensate.
Thus we define the impurity energy $\delta F$ in Eq.~\eqref{eq:impurity_energy}
by subtracting the free energy of light quarks, $-\Tr \, \ln S_{0}(x)^{-1}$, from the free energy~\eqref{eq:free_energy}.

The trace of the second term in Eq.~\eqref{eq:eff_pot_Kondo_single_GL} is calculated as follows.
For a while, we omit the trace for the color space for brevity.
First of all, we find that the terms with $k$ odd vanish as it can be checked directly.
Thus, we consider only the terms with $k$ even.
As a demonstration, we calculate the $k=2$ term:
\begin{widetext}
\begin{eqnarray}
   \Tr
   \Bigl( S_{0}(x) \hat{\Delta}(x) \Bigr)^{2}
&=&
   2 \, \Tr
   \biggl(
         \frac{1+\gamma_{4}}{2}
         \frac{1}{ \partial_{\tau} + \lambda } \Delta(\tau)
         \frac{1}{\gamma\partial+m-\mu\gamma_{4}} \bar{\Delta}(\tau)
         \delta(\vec{x})
   \biggr)
\nonumber \\ 
&\approx&
   2 \, \tr
   \int \drm^{4}x \,
   \langle x |
   \frac{1+\gamma_{4}}{2} \frac{1}{ \partial_{\tau} + \lambda } \Delta(\tau)
   \frac{1}{\gamma\partial+m-\mu\gamma_{4}} \bar{\Delta}(\tau)
   \delta(\vec{x})
   | x \rangle
\nonumber \\ 
&=&
   2 \, \tr
   \int \drm^{4}x
   \int \frac{\drm^{4}p}{(2\pi)^{4}} \,
   \langle x |
   \frac{1+\gamma_{4}}{2} \frac{1}{ \partial_{\tau} + \lambda } \Delta(\tau)
   \frac{1}{\gamma\partial+m-\mu\gamma_{4}}
   \bar{\Delta}(\tau)
   \delta(\vec{x})
   | p \rangle \langle p | x \rangle
\nonumber \\ 
&=&
   2
   \int \drm\tau
   \int \frac{\drm^{4}p}{(2\pi)^{4}}
   \frac{ ip_{4} + \lambda }{ p_{4}^{2} + \lambda^{2} }
   \frac{1}{(p_{4}-\mu)^{2}+\vec{p}^{2}+m^{2}}
   \tr
   \biggl(
         \frac{1+\gamma_{4}}{2}
         \Delta(\tau)
         \bigl(ip\gamma+m-\mu\gamma_{4}\bigr)
         \bar{\Delta}(\tau)
   \biggr)
\nonumber \\[0em] 
&=&
   2
   \int \drm\tau \,
   T\sum_{n}
   \Biggl(
   \frac{1}{i\omega_{n}-\lambda}
   \int \frac{\drm^{3}\vec{p}}{(2\pi)^{3}}
   \frac{1}{(i\omega_{n}-\mu)^{2}-E_{\vec{p}}^{2}}
   \biggl(
         \Bigl(
               (1+\alpha_{\vec{p}}^{2}) (i\omega_{n}-\mu)
            + (1-\alpha_{\vec{p}}^{2}) m
             - 2 \alpha_{\vec{p}} |\vec{p}|
         \Bigr)
         \Phi \Phi^{\dag}
         \nonumber \\ && 
      + \Bigl(
               (1+\alpha_{5\vec{p}}^{2}) (i\omega_{n}-\mu)
             - (1-\alpha_{5\vec{p}}^{2}) m
             - 2 \alpha_{5\vec{p}} |\vec{p}|
         \Bigr)
         \Phi_{5} \Phi_{5}^{\dag}
   \biggr)
   \Biggr),
\end{eqnarray}
\end{widetext}
where the first trace ($\mathrm{Tr}$) means the diagonal summation both in the Dirac space and in the Euclidean space,
and the second trace ($\tr$) means the diagonal summation in the Dirac space.
We have used the 
 approximation that the change of the fields over spacetime is sufficiently smooth in the low-energy limit and neglected the derivative terms.
In the last transformation, we have used the relations
$\vec{\Phi} = \alpha_{\vec{p}} \hat{\vec{p}} \, \Phi$ and $\vec{\Phi}_{5} = i \alpha_{5\vec{p}} \hat{\vec{p}} \, \Phi_{5}$
with $\alpha_{\vec{p}}=-|\vec{p}|/(E_{\vec{p}}+m)$, $\alpha_{5\vec{p}}=-|\vec{p}|/(E_{\vec{p}}-m)$, and $\hat{\vec{p}}=\vec{p}/|\vec{p}|$
in Eqs.~\eqref{eq:DeltaPpm_1} and \eqref{eq:DeltaPpm_2}
 and changed the $p_{4}$ integral to the sum over the Matsubara frequency $\omega_{n}=(2n+1)\pi/\beta$ ($n \in \mathbb{Z}$) with   the inverse temperature $\beta=1/T$.

Similarly, for general $k$ (even), we obtain their sum:
\begin{widetext}
\begin{eqnarray}
 - \sum_{k\ge1} \frac{(-1)^{k+1}}{k}
   \Tr
   \Bigl( S_{0}(x) \hat{\Delta}(x) \Bigr)^{k}
&=&
 - \tr
   \int \drm\tau \,
   T\sum_{n}
   \ln
   \Biggl(
   1
 - \frac{1}{i\omega_{n}-\lambda}
   \int \frac{\drm^{3}\vec{p}}{(2\pi)^{3}}
   \frac{1}{(i\omega_{n}-\mu)^{2}-E_{\vec{p}}^{2}}
   \nonumber \\ && \hspace{0em} \times 
   \biggl(
         \Bigl(
               (1+\alpha_{\vec{p}}^{2}) (i\omega_{n}-\mu)
            + (1-\alpha_{\vec{p}}^{2}) m
             - 2 \alpha_{\vec{p}} |\vec{p}|
         \Bigr)
         \Phi \Phi^{\dag}
         \nonumber \\ && 
      + \Bigl(
               (1+\alpha_{5\vec{p}}^{2}) (i\omega_{n}-\mu)
             - (1-\alpha_{5\vec{p}}^{2}) m
             - 2 \alpha_{5\vec{p}} |\vec{p}|
         \Bigr)
         \Phi_{5} \Phi_{5}^{\dag}
   \biggr)
   \Biggr).
\end{eqnarray}
\end{widetext}
The summation over the Matsubara frequencies is taken by using the formula
\begin{eqnarray}
   T\sum_{n} F(i\omega_{n})
&=&
 - \frac{1}{2\pi i}
   \int_{-\infty}^{\infty} \drm p_{0}
   \bigl(F_{+}(p_{0})-F_{-}(p_{0})\bigr)
   \nonumber \\ && \times 
   \frac{1}{e^{\beta p_{0}}+1},
\end{eqnarray}
with $F_{\pm}(p_{0})=F(p_{0}\pm i\varepsilon)$
for an analytic function $F$ with the branch cut on the real axis and no poles elsewhere.
Note that $p_{0}$ is a real number, and $\varepsilon$ is an infinitely small and positive number.

From the above calculations, we obtain
the concrete form of the impurity energies shown in Eqs.~\eqref{eq:eff_pot_Lambda_0}, \eqref{eq:eff_pot_Lambda_p} and \eqref{eq:eff_pot_Lambda_ap}.
In Eq.~\eqref{eq:eff_pot_Lambda_0}, ${\cal I}_{N_{\pm}}(\omega)$ are defined as
\begin{eqnarray}
   {\cal I}_{N_{+}}(\omega)
&=&
 - (\omega + m) \, {\cal I}_{N}(\omega)
+ (-\omega - m) \, {\cal I}_{N}(-\omega),
\label{eq:IN+_def} \\ 
   {\cal I}_{N_{-}}(\omega)
&=&
 - (\omega - m) \, {\cal I}_{N}(\omega)
+ (-\omega + m) \, {\cal I}_{N}(-\omega),
\label{eq:IN-_def} 
\end{eqnarray}
where
\begin{eqnarray}
   {\cal I}_{N}(\omega)
&=&
   \int_{m}^{\Lambda_{m}} \drm \xi \, \sqrt{\xi^{2}-m^{2}} \,
   \frac{1}{\xi-\omega}
\nonumber \\[0.5em] 
&=&
   \Lambda
 - \omega
   \ln\biggl(\frac{m}{\Lambda+\Lambda_{m}}\biggr)
+ \theta(|\omega|-m)
   \sqrt{\omega^{2}-m^{2}}
   \nonumber \\ && \times 
   \ln
   \Biggl|
         \frac{m\bigl(-\omega+\Lambda_{m}\bigr)}
         {-m^{2}+\Lambda\sqrt{\omega^{2}-m^{2}}+\omega\Lambda_{m}}
   \Biggr|
   \nonumber \\ && 
 - \theta(m-|\omega|)
   \sqrt{m^{2}-\omega^{2}} \,
   \Biggl(
         \atan
         \Biggl(
               \frac{\Lambda\sqrt{m^{2}-\omega^{2}}}{m^{2}-\omega\Lambda_{m}}
         \Biggr)
         \nonumber \\ && 
      + \pi \,
         \theta
         \biggl(
               \omega-\frac{m^{2}}{\Lambda_{m}}
         \biggr)
   \Biggr).
\end{eqnarray}
In Eq.~\eqref{eq:eff_pot_Lambda_p}, ${\cal I}_{P}(\omega)$ is defined as
\begin{eqnarray}
   {\cal I}_{P}(\omega)
&=&
   \mathrm{P.V.}
   \int_{m}^{\Lambda_{m}} \drm \xi \,
   \frac{\xi\sqrt{\xi^{2}-m^{2}}}{\omega-\xi}
\nonumber \\[0.5em] 
&=&
 - \Lambda \omega
 - \frac{\Lambda\Lambda_{m}}{2}
+ \biggl(\omega^{2}-\frac{m^{2}}{2}\biggr)
   \ln\biggl(\frac{m}{\Lambda+\Lambda_{m}}\biggr)
   \nonumber \\ && 
+ \omega \sqrt{\omega^{2}-m^{2}}
   \ln\biggl(
   \frac{\Lambda\sqrt{\omega^{2}-m^{2}}+\omega\Lambda_{m}-m^{2}}{m\bigl(\Lambda_{m}-\omega\bigr)}
   \biggr),
\nonumber \\ 
\label{eq:IP_def}
\end{eqnarray}
with $\Lambda_{m}=\sqrt{\Lambda^{2}+m^{2}}$,
where $\xi$ is the energy variable in the integral, and $\mathrm{P.V.}$ stands for the Cauchy's principal value.
In Eq.~\eqref{eq:eff_pot_Lambda_ap}, ${\cal I}_{AP}(\omega)$ is defined as
\begin{eqnarray}
   {\cal I}_{AP}(\omega)
&=&
   \mathrm{P.V.}
   \int_{m}^{\Lambda_{m}} \drm \xi \,
   \frac{\xi\sqrt{\xi^{2}-m^{2}}}{\omega+\xi}
\nonumber \\[0.5em] 
&=&
 - \Lambda \omega
+ \frac{\Lambda\Lambda_{m}}{2}
 - \biggl(\omega^{2}-\frac{m^{2}}{2}\biggr)
   \ln\biggl(\frac{m}{\Lambda+\Lambda_{m}}\biggr)
   \nonumber \\ && 
+ \omega \sqrt{\omega^{2}-m^{2}}
   \ln\biggl(
   \frac{\Lambda\sqrt{\omega^{2}-m^{2}}-\omega\Lambda_{m}-m^{2}}{m\bigl(\Lambda_{m}+\omega\bigr)}
   \biggr).
\nonumber \\ 
\label{eq:IAP_def}
\end{eqnarray}

We calculate the third term in the right-hand side of Eq.~\eqref{eq:eff_pot_Kondo_single_GL} in the momentum space as
\begin{widetext}
\begin{eqnarray}
   \frac{1}{G} \int \drm\tau \,
   \bigl( \Phi^{\dag}\Phi + \Phi_{5}^{\dag}\Phi_{5} + \vec{\Phi}^{\dag}\vec{\Phi} + \vec{\Phi}_{5}^{\dag}\vec{\Phi}_{5} \bigr)
&=&
   \frac{1}{G} \int \drm\tau
   \int \drm^{3}\vec{x} \,
   \delta(\vec{x})
   \bigl( \Phi^{\dag}\Phi + \Phi_{5}^{\dag}\Phi_{5} + \vec{\Phi}^{\dag}\vec{\Phi} + \vec{\Phi}_{5}^{\dag}\vec{\Phi}_{5} \bigr)
\nonumber \\[0em] 
&=&
   \frac{1}{G} \int \drm\tau
   \int \drm^{3}\vec{x}
   \int \frac{\drm^{3}\vec{p}}{(2\pi)^{3}} e^{i\vec{p}\cdot\vec{x}} \,
   \Bigl( \bigl(1+\alpha_{\vec{p}}^{2}\bigr) \Phi^{\dag}\Phi + \bigl(1+\alpha_{5\vec{p}}^{2}\bigr) \Phi_{5}^{\dag}\Phi_{5} \Bigr)
\nonumber \\[0em] 
&=&
   \frac{1}{G} \int \drm\tau
   \Bigl( \bigl(1+\alpha_{\vec{p}}^{2}\bigr) \Phi^{\dag}\Phi + \bigl(1+\alpha_{5\vec{p}}^{2}\bigr) \Phi_{5}^{\dag}\Phi_{5} \Bigr)
   \Bigr|_{\vec{p}=\vec{0}}.
\end{eqnarray}
\end{widetext}
where the last equation can be further modified by substituting the QCD Kondo condensates.
Recovering the color factor in the trace, finally, we obtain the free energy~\eqref{eq:free_energy}.

\section{Simple solution of $\delta F_{P_{\pm}}$ at large $\mu$}
\label{sec:simple_solution}

At large chemical potential ($\mu \gg m$), we can obtain a simple form of the solution in the Kondo resonance of the $P_{\pm}$ condensate (see also Ref.~\cite{Yasui:2016yet}).
In this limit, the energy integral in Eq.~\eqref{eq:eff_pot_Lambda_p_2} is analytically performed and the impurity energy is expressed in approximation as
\begin{eqnarray}
   \delta F_{P_{\pm}}
&\approx&
   \frac{2N_{c}}{\pi}
   \Biggl(
       - \lambda \, \atan \biggl(\frac{\lambda}{\Delta_{P}}\biggr)
      + \lambda \, \atan \biggl(\frac{\mu+\lambda}{\Delta_{P}}\biggr)
         \nonumber \\ && 
      + \frac{1}{2} \Delta_{P}
         \ln\biggl(\frac{\Delta_{P}^{2}+\lambda^{2}}{\Delta_{P}^{2}+(\mu+\lambda)^{2}}\biggr)
   \Biggr)
+ \frac{1}{\kappa G} \Delta_{P}
   \nonumber \\ && \hspace{0em} 
+ 2N_{c} \lambda \theta(-\lambda)
 - \lambda,
\label{eq:free_energy_P_approximation}
\end{eqnarray}
with $\kappa=\pi \mu \sqrt{\mu^{2}-m^{2}}/(2\pi^{2})$, where we assumed the large chemical potential and the small value for $\Delta_{P}$.
From the stationary conditions, $\partial \delta F_{P_{\pm}}/\partial \Delta_{P}=\partial \delta F_{P_{\pm}}/\partial \lambda=0$, we obtain the QCD Kondo condensate and the Lagrange multiplier,
\begin{eqnarray}
   \Delta_{P}
&\approx&
   \frac{\pi}{2N_{c}}
   \mu
   \exp
   \biggl(
      -\frac{2\pi^{2}}{2N_{c}\mu \sqrt{\mu^{2}-m^{2}}G}
   \biggr),
\label{eq:gap_P_approximation}
\\[0.5em] 
   \lambda
&\approx&
   \mu
   \exp
   \biggl(
      -\frac{2\pi^{2}}{2N_{c}\mu \sqrt{\mu^{2}-m^{2}}G}
   \biggr),
\label{eq:lambda_P_approximation}
\end{eqnarray}
in the large $N_{c}$ approximation.
Notice that the sign of $\lambda$ is positive as long as $\mu>0$, which means that the QCD Kondo resonance emerges around the energy $\lambda$ above the Fermi surface, as shown by Eq.~\eqref{eq:spectral_function_P}.
Substituting $\Delta_{P}$ and $\lambda$ into Eq.~\eqref{eq:free_energy_P_approximation}, the impurity energy becomes
\begin{eqnarray}
   \delta F_{P_{\pm}}
&\approx&
 - \mu
   \exp\biggl(
      -\frac{2\pi^{2}}{2N_{c}\mu \sqrt{\mu^{2}-m^{2}}G}
   \biggr) < 0.
\end{eqnarray}
The sign of the impurity energy is negative.
Thus, we confirm that the $P_{\pm}$ condensate is realized for any (nonzero) coupling constant $G$ as long as the interaction is attractive.

In the large $N_{c}$ limit, where $N_{c}G$ is a finite value (the 't~Hooft limit), $\Delta_{P}$ approaches zero asymptotically from Eq.~\eqref{eq:gap_P_approximation}, while $\lambda$ remains finite  from Eq.~\eqref{eq:lambda_P_approximation}: the QCD Kondo resonance becomes sharp around the energy position $\lambda$.
We confirm, however, that the QCD Kondo effect does not disappear, because $\delta F_{P_{\pm}}$ is still a negative nonzero value.

\bibliography{reference_2}

\end{document}